\documentclass[fleqn]{2023SCGE}
\usepackage{graphicx}

\newcommand{\msun}{M$_{\odot}$\ }
\newcommand{\rsun}{R$_{\odot}$\ }
\newcommand{\lsun}{L$_{\odot}$\ }
\newcommand{\lam}{\lambda}
\newcommand{\Ni}{$^{56}$Ni\ }

\begin{document}

\ensubject{subject}

\ArticleType{Article}
\SpecialTopic{SPECIAL TOPIC: }
\Year{2023}
\Month{xx}
\Vol{xx}
\No{xx}
\DOI{??}
\ArtNo{000000}
\ReceiveDate{October 13, 2023}
\AcceptDate{October 30, 2023}

\title{The Dusty and Extremely Red Progenitor of the Type II \\ Supernova 2023ixf in Messier 101}

\author[1]{Danfeng Xiang}{}
\author[1]{Jun Mo}{}
\author[2]{Lingzhi Wang}{}
\author[1,3]{Xiaofeng Wang}{wang\_xf@mail.tsinghua.edu.cn}
\author[4,5,6]{Jujia Zhang}{}
\author[4,5]{Han Lin}{}
\author[7]{Lifan Wang}{}

\AuthorMark{Xiang D F}

\AuthorCitation{Xiang D F, Mo J, Wang L Z, et al}

\address[1]{Department of Physics and Tsinghua Center for Astrophysics, Tsinghua University, Haidian District, Beijing 100084, China}
\address[2]{South America Center for Astronomy, National Astronomical Observatories, Chinese Academy of Sciences, Beijing 100101, China}
\address[3]{Beijing Planetarium, Beijing Academy of Sciences and Technology, Beijing 100044, China}
\address[4]{Yunnan Observatories, Chinese Academy of Sciences, Kunming, 650216, China}
\address[5]{Key Laboratory for the Structure and Evolution of Celestial Objects, Chinese Academy of Sciences, Kunming, 650216, China}
\address[6]{International Centre of Supernovae, Yunnan Key Laboratory, Kunming 650216, China}
\address[7]{Mitchell Institute for Fundamental Physics and Astronomy, Texas A\&M University, College Station, TX 77843, USA}

%
\abstract{Stars with initial masses in the range of 8--25 solar masses are thought to end their lives as hydrogen-rich supernovae (SNe II). Based on the pre-explosion images of Hubble Space Telescope (\textit{HST}) and \textit{Spitzer} Space Telescope, we place tight constraints on the progenitor candidate of type IIP SN 2023ixf in Messier 101. Fitting of the spectral energy distribution (SED) of its progenitor with dusty stellar spectral models results in an estimation of the effective temperature as 3091$^{+422}_{-258}$~K. The luminosity is estimated as log($L/$L$_{\odot}$)$\sim4.83$, consistent with a red supergiant (RSG) star with an initial mass of 12$^{+2}_{-1}$~M$_{\odot}$. The derived mass loss rate (6--9$\times10^{-6}$~\msun yr$^{-1}$) is much lower than that inferred from the flash spectroscopy of the SN, suggesting that the progenitor experienced a sudden increase in mass loss when approaching the final explosion. In the infrared bands, significant deviation from the range of regular RSGs in the color-magnitude diagram and period-luminosity space of the progenitor star indicates enhanced mass loss and dust formation. Combining with new evidence of polarization at the early phases of SN 2023ixf, such a violent mass loss is likely a result of binary interaction.}



\maketitle


\begin{multicols}{2}
\section{Introduction} \label{sec:intro}

Type II Supernovae (SNe II) are thought to be produced by core-collapse of red supergiants (RSGs), which have initial masses of 8--25 \msun \cite{2003ApJ...591..288H}. These stars retain most of their hydrogen envelopes before core collapse, producing supernovae with prominent hydrogen lines. 
\Authorfootnote 
And the light curves of SNe IIP display plateau features lasting up to about 100 days after a rapid rise. 
With pre-discovery images, progenitors have been identified for dozens of SNe II \cite{2015PASA...32...16S} and more recently for SN~2017eaw 
\cite{2019ApJ...875..136V,2019MNRAS.485.1990R,2018MNRAS.481.2536K} and SN~2022acko\cite{2023MNRAS.524.2186V}. These observations have confirmed the connections between RSGs and SNe II, with exceptions of the famous SN~1987A and 87A-like objects whose progenitor stars are believed to be blue supergiants (BSGs) \cite{1988ApJ...324..466W,2023MNRAS.520.2965X}. Special mechanisms, such as semi-convection and binary interaction, were required to produce a BSG supernova progenitor \cite{1987ApJ...319..136A,1987Natur.327..597H,1988ApJ...331..388S,1991A&A...252..669L,1989Natur.338..401P,2007AIPC..937..125P}.
Moreover, stars with initial masses in the range of 8--11 \msun are thought to form a degenerate O+Ne+Mg core instead of an iron core, and they tend to become super asymptotic giant branch (sAGB) stars at the end of their lives \cite{1984ApJ...277..791N,1987ApJ...322..206N,1992ApJ...396..649T,2004ApJ...612.1044P,
2015MNRAS.451.2123T}. The electron-capture onto Ne and Mg would accelerate the contraction of stellar core, which will result in an electron-captured supernova (ECSN) that may help explain some subluminous SNe IIP \cite{2006A&A...450..345K,2008ApJ...675..614P,2015MNRAS.446.2599D,2022MNRAS.509.2013Z,2020MNRAS.498...84Z,2021NatAs...5..903H}.

On the other hand, the lack of discovery of RSGs with initial masses $>$17~\msun as progenitors of SNe IIP challenges current theories of massive stellar evolution, i.e., the ``red supergiant problem'' \cite{2009MNRAS.395.1409S}. It is suggested that most massive stars above 20~\msun may collapse quietly to black holes so that the explosions are too faint to have been detected \cite{2009ARA&A..47...63S}. On the other hand, the problem can probably also be explained by the dust surrounding the progenitor star which can cause underestimates of the luminosities of the progenitor stars \cite{2012ApJ...759...20K,2018MNRAS.481.2536K}. The mass lost by the progenitor star tends to form circumstellar (CS) dust obscuring the star light severely in visual bands. In this case, interaction signatures, typically narrow emission lines, are also expected to be observed in the early spectra of some SNe II. The dust is quickly destroyed by the emission of the explosion and the remaining gas collides with the expanding SN ejecta, making the explosion appear more energetic. In addition, it is thought that mass loss shortly prior to explosion happens to a large fraction of SN II progenitors \cite{2017MNRAS.470.1642F,2020ApJ...889...86W,2021ApJ...912...46B,2023MNRAS.523.1474R}, evidenced by transient emission lines disappearing shortly after the supernova explosion (e.g., \cite{2002ApJ...572..932P,2007ApJ...666.1093Q,2019MNRAS.485.1990R}).

On 2023 May 19.728, 2023, the amateur astronomer Koichi Itagaki discovered a new possible supernova (SN) in the outskirt of Messier 101 which is a nearby face-on spiral galaxy at a distance of 6.85$\pm$0.13 Mpc \cite{2022ApJ...934L...7R}. This stellar explosion event, later named as SN~2023ixf, was soon confirmed to be a hydrogen-rich (type II) supernova with strong flash ionization lines of H, He, C, and N in the early spectra\footnote{\url{https://www.wis-tns.org/object/2023ixf}}. 
The SN site has subsolar metallicity of 12+log(O/H) = 8.45$\pm$0.03 (i.e. [Fe/H] $\approx-$~0.24 \cite{2016ApJ...830....4C}). A similar metallicity 12+log(O/H) = 8.37$\pm$0.18 was found by measuring the nebular emission lines of the nearby HII regions \cite{2023ApJ...955L..15N}.
A small reddening of $E(B-V)$ = 0.03 mag can be inferred for the host galaxy from the weak Na~I~D lines in the high resolution spectra of SN~2023ixf \cite{2023arXiv230901998Z,2023TNSAN.160....1L,2023arXiv230607964S,2023ApJ...954L..12T}. Including the galactic reddening of $\sim$0.01~mag, the total reddening to SN~2023ixf is given as $E(B-V)=0.04$ mag.

Follow-up observations indicate that SN~2023ixf is a luminous type IIP supernova (Li et al. in Prep.). 
Interestingly, the very early time (t$<$0.3 day) color of this SN is quite red and it then turns blue quickly, indicating thick dust surrounding the pre-exploding star which is destroyed shortly (within 0.3 day) after the shock breakout. Meanwhile, the narrow emission lines of ionized He, C, N and hydrogen diminish quickly within one week after the explosion \cite{2023ApJ...955L...8H,2023PASJ...75L..27Y,2023ApJ...956L...5B,2023arXiv230901998Z}, indicating that the supernova may have strong interactions with some H-rich circumstellar matter (CSM) which is located very close to the progenitor star. And the CSM is significantly aspherical according to the asymmetric structure of the emission lines \cite{2023ApJ...956...46S} and spectropolarimetry observations \cite{2023ApJ...955L..37V}.

Immediately after the discovery of SN 2023ixf, numerous papers on the observations and properties analysis of its progenitor candidate emerged \cite{2023ApJ...952L..23K,2023ApJ...952L..30J,2023arXiv230610783S,2023ApJ...955L..15N,2023arXiv230910022Q,2023arXiv230814844V}. These studies, which used similar data set and methods, reached similar results that the progenitor star was an RSG with a thick dust shell, but with a large range of initial mass (11--24~M$_{\odot}$), mainly due to different estimates of the stellar luminosity. 
Spectral energy distribution (SED) from optical to near/mid-infrared (NIR/MIR) bands was used to fit stellar spectral models to derive the parameters of the progenitor candidate of SN 2023ixf except in ref.~\cite{2023arXiv230610783S}.
Ref.~\cite{2023arXiv230610783S} obtained the highest luminosity (log$L/\mathrm{L_{\odot}}]\sim5.37$), hence the most massive progenitor, using the period-luminosity relation of RSGs. While as we will show, the progenitor star does not belong to normal RSGs in the $P-L$ diagrams. This method would result in overestimate of the luminosity.
Ref.~\cite{2023ApJ...952L..23K} was the first to identify the progenitor as an RSG, but they used a single phase $K$-band magnitude which was 2 magnitude fainter than other studies (and this work), perhaps mainly due to the variability in NIR bands of the star unconvered later by ref.~\cite{2023arXiv230610783S}.
Fitting of the stellar SED using only several fixed efficient temperatures in the range of 3400--4000 K has found higher luminosity could be obtained with higher presumed temperature \cite{2023ApJ...955L..15N}.
Among these results, ref.~\cite{2023arXiv230814844V} found similar results to ours, although they used different stellar models.
While ref.~\cite{2023ApJ...952L..30J} and ref.~\cite{2023arXiv230910022Q} both suggested a higher initial mass of $\gtrsim$17~M$_{\odot}$.


In this paper, we present detailed analysis of the multiband photometric results from optical to mid-infrared (MIR) bands based on the archived pre-explosion images from space and ground-based telescopes. Our analysis is updated with new observational results of the SN itself. These data allow us to put stringent constraints on the progenitor star and the circumstellar dust around it, which is important to the study of final-stage evolution of SNe II progenitors. 

\section{Data} \label{sec:methods}
\subsection{Ground-based near-infrared data}\label{sec:lit-data}
The near-infrared (NIR) photometry of the progenitor of SN 2023ixf was obtained with the NEWFIRM infrared camera mounted on the Gemini North telescope and the Wide Field Camera (WFCAM) mounted on the 3.8-m United Kingdom Infrared Telescope (UKIRT) \cite{2023ApJ...952L..23K,2023ApJ...952L..30J,2023arXiv230610783S}. The progenitor is detected in the \textit{JHK}-bands and exhibits significant variability with amplitudes of about 0.5 mag \cite{2023arXiv230610783S}. In our analysis, we directly use the average magnitudes of their results in our analysis in Section \ref{sec:properties}, which are $m_J$ = 20.63$\pm$0.34 mag, $m_H$ = 19.63$\pm$0.38 mag, and $m_K$ = 18.73$\pm$0.22 mag, respectively.
\begin{figure*}[htbp!]
	\centering
	\includegraphics[width=\linewidth]{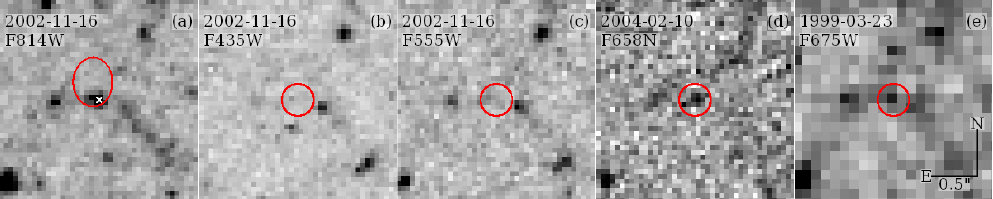}
	\caption{(a): The region around the site of \mbox{SN~2023ixf} showing on the pre-discovery \textit{HST} F814W image. The white cross marks the location of the progenitor star candidate. The center of red ellipses shows the SN position, with the radius of the ellipse showing the size of the error (1-$\sigma$). (b)$\sim$(e): the pre-explosion \textit{HST} images centering around the progenitor candidate in the F435W, F555W, F658N, and F675W bands, respectively, with the center of the circles marking the center positions of the identified progenitor star. All of the above images are aligned. North is up and east is to the left.\label{fig:hstimg}}
\end{figure*}
\begin{table}[H]
	\footnotesize
	\caption{Photometry results of the pre-explosion \textit{HST} images at the site of \mbox{SN~2023ixf}. All magnitudes are in the Vega system. Detection limits are given at 5-$\sigma$.\label{tab:hst-obs}}
	\tabcolsep 4pt 
	\centering
	\begin{tabular}{ccccc}
	\toprule
	Obs. date &Instrument &Filter  &mag &1-$\sigma$ error\\
	\hline
	1999-03-23 & WFPC2/WFC   & F675W & 24.419  & 0.191  \\
	2002-11-16 & ACS/WFC     & F435W & 28.957  & 1.193  \\
	2002-11-16 & ACS/WFC     & F555W & 28.599  & 1.233  \\
	2002-11-16 & ACS/WFC     & F814W & 24.266  & 0.045  \\
	2004-02-10 & ACS/WFC     & F658N & 24.618  & 0.179  \\
	1999-03-23 & WFPC2/WFC   & F547M & $>$25.8 & \\
	1999-03-23 & WFPC2/WFC   & F656N & $>$21.7 & \\
	1999-03-23 & WFPC2/WFC   & F675W & $>$25.6 & \\
	1999-06-17 & WFPC2/WFC   & F547M & $>$25.7 & \\
	1999-06-17 & WFPC2/WFC   & F656N & $>$21.9 & \\
	2003-08-27 & WFPC2/WFC   & F336W & $>$23.7 & \\
	2014-03-19 & WFC3/UVIS   & F502N & $>$24.8 & \\
	2014-03-19 & WFC3/UVIS   & F673N & $>$24.6 & \\
	2018-03-30 & ACS/WFC     & F435W & $>$28.8 & \\
	2018-03-30 & ACS/WFC     & F658N & $>$25.7 & \\
\bottomrule
\end{tabular}
\end{table}

\subsection{Pre-explosion optical and MIR photometry from space-based telescopes}\label{sec:HST_phot}
The Hubble Space Telescope (\textit{HST}) observed the SN site in various bands during the period from 1999 to 2018. A point source can be clearly seen in the F658N and F814W-band images at the position coincident with the SN site. The pre-discovery \textit{HST} images around the SN position are shown in Figure \ref{fig:hstimg}. 
Details of the \textit{HST} images and data reduction is presented in Appendix \ref{sec:HST} and the photometric results are presented in Table~\ref{tab:hst-obs}.
We note that the source detected on the F547M and F675W-bands images on Mar. 23rd, 1999 is marked as hot spots by DOLPHOT. In the F658N-band, we found that the progenitor has darkened by $\geq$1.1 mag from 2004 to 2018. The narrow band F658N traces the wavelength region of H$\alpha$ line, not the continuum flux density. So we do not include the results of F547M, F675W and F658N in the SED fitting.
Our phototmetric results of $HST$ images are consistent with those of refs.~\cite{2023arXiv230814844V,2023ApJ...955L..15N}, while ref.~\cite{2023arXiv230814844V} used F675W, F658N, F673N, F814W in their SED fitting. Ref.~\cite{2023ApJ...952L..30J} used the same $HST$ magnitudes as in ref.~\cite{2023ApJ...952L..23K} which was $\sim$0.6 magnitude fainter than our result in F814-band.

\begin{figure}[H]
	\centering
	\includegraphics[width=1.0\columnwidth]{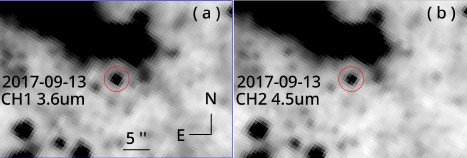}
	\caption{The pre-discovery \textit{Spitzer}/IRAC $CH1$ (panel a) and $CH2$-band (panel b) images around the SN~2023ixf site, taken on Sept. 13, 2017 are marked with red circles with a radius of 2.5 arcsec.\label{fig:spitzer-img}}
\end{figure}
\begin{figure}[H]
	\centering
	\includegraphics[width=1.0\linewidth]{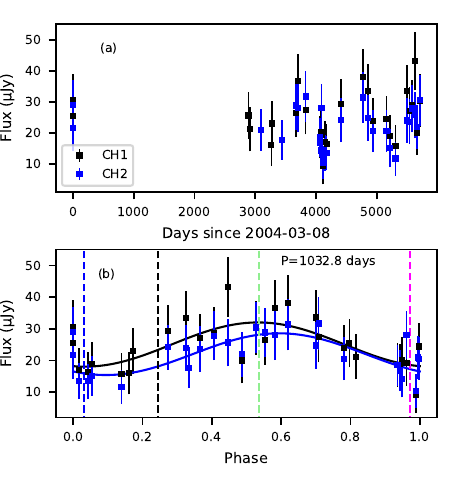}
	\caption{(a): \textit{Spitzer}/IRAC $CH1$- and $CH2$-band light curves of the SN~2023ixf progenitor; (b): the phased light curves folded by a period of 9997.5 days. In the lower panel, the corresponding phases of the dates of \textit{HST} images, 1999-03-23, 2002-11-16, 2004-04-10, and 2018-03-30 are labeled as black, green, blue, and magenta dashed lines, respectively.  \label{fig:spitzer-lc}}
\end{figure}

The SN~2023ixf field in M101 was observed with the \textit{Spitzer} Infrared Array Camera (IRAC) before its explosion by several programs covering the phases from 2004 to 2019. A point source is clearly detected at a $2\sigma$ threshold in $CH1$ ($3.6\mu$m) and $CH2$ ($4.5\mu$m) bands during the period from 2004 to 2019, as shown in Figure~\ref{fig:spitzer-img}, while there is no detection in $CH3$ ($5.8\mu$m) and $CH4$ ($8.0\mu$m) bands in 2004. Aperture photometry was performed on the pre-explosion images of the SN field, and the AB magnitudes and fluxes of the progenitor star measured in $CH1$ and $CH2$ bands are obtained (described in Appendix \ref{sec:Spitzer} and displayed in Table \ref{tab:spitzer-phot}). The progenitor star candidate is measured to have a median flux of $24.43\pm7.71 \ \mu$Jy in $CH1$ band and $21.97\pm6.33\ \mu$Jy in $CH2$ band, respectively, with the corresponding AB magnitudes being $20.43\pm0.36$~mag and $20.55\pm0.34$~mag. While the corresponding Vega magnitudes are $17.65\pm0.36$~mag in $CH1$ band and $17.28\pm0.34$~mag in the CH2 band, respectively, which are consistent with the measurements by other researchers \cite{2023ATel16042....1S}. 
The light curves in $CH1$ and $CH2$ are displayed in Figure \ref{fig:spitzer-lc}, exhibiting periodic fluctuations, which are further studied in Section \ref{sec:spitzer-variation}.

\section{Results}

\subsection{Infrared variability and colors of the progenitor} \label{sec:spitzer-variation}
Pre-explosion \textit{Spitzer}/IRAC $CH1$- and $CH2$-band light curves of the progenitor display a periodic variability indicating pulsational activity which is commonly seen in RSGs and AGB stars.
Stellar variability and periodicity are performed on the \textit{Spitzer}/IRAC $CH1$- and $CH2$-band light curves in order to study the MIR evolution of the progenitor. The MIR light curves span 5709 days and show fluctuations, which are also mentioned in previous studies \cite{2023ATel16042....1S,2023ApJ...952L..23K,2023ApJ...952L..30J,2023arXiv230610783S}. 
We searched for periodic variability using the Lomb-Scargle method \cite{1976Ap&SS..39..447L,1982ApJ...263..835S} through \texttt{VARTOOLS} program \cite{Hartman2016}. A long period of 967.7 days with the highest signal noise ratio S/N = 9.2 can be detected in $CH2$-band light curve between 100 and 2000 days with a bin size of 0.1 day. Similarly, a period of 1097.9 days can be detected in the $CH1$ band with S/N = 4.2. 
We then phased the $CH1$- and $CH2$-band light curves by the mean long period of 1032.8 days, as shown in Figure~\ref{fig:spitzer-lc}. At the same time, three epochs on 2002-11-16, 2004-2-10, and 2018-3-30 observed by the \textit{HST} were overplotted on the phased light curves. 
In order to combine the \textit{Spitzer} MIR flux with those of optical (by \textit{HST}) and NIR data, we put the date of the \textit{HST} images (2002-11-16) on the periodic IRAC-$CH1/CH2$ light curves.
Then the flux in IRAC $CH1$ and $CH2$ bands are estimated to be 31.33$\pm$8.94~$\mu$Jy and 26.99$\pm$3.29~$\mu$Jy and are used in the analysis in Section \ref{sec:sed-fit}.
We note that the \textit{HST} epoch is located near the peak of the pulsation cycle, thus the fluxes in $CH1$- and $CH2$- bands are higher than in other works since either average values \cite{2023ApJ...952L..23K,2023arXiv230814844V,2023ApJ...955L..15N} or phased values near the cycle bottom \cite{2023ApJ...952L..30J} were adopted.

\begin{figure*}[ht!]
	\centering
	\includegraphics[width=0.9\linewidth]{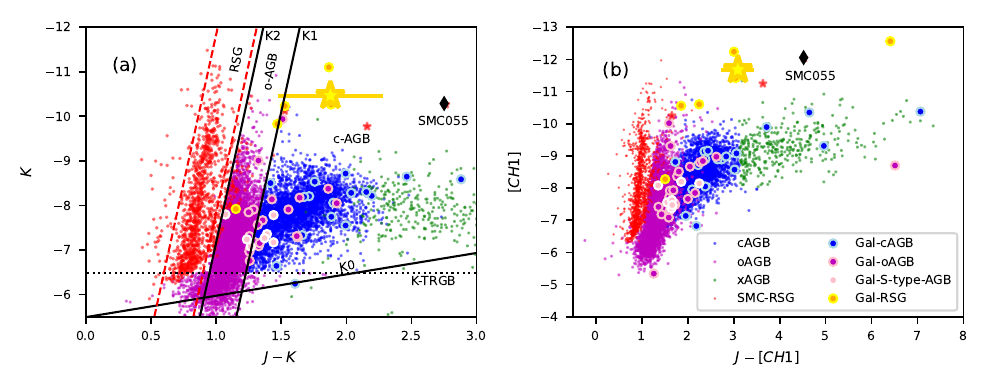}
	\caption{Location of the progenitor of SN 2023ixf (golden star) in the $K$ vs. $J-K$ (panel a) and $[CH1]$ vs. $J-[CH1]$ (panel b) Color-Magnitude diagrams. Criteria (K0, K1, K2) correspond to the NIR color for cAGBs, oAGBs and RSGs \cite{2006A&A...448...77C} are plotted as solid lines. RSGs with $J-K>1.5$~mag are marked as red stars. For comparison, AGB \cite{2010ApJ...723.1195R} and RSG \cite{2020A&A...639A.116Y} samples are plotted as small colored dots. Also plotted as filled dots are Galactic AGB stars and RSGs \cite{2015MNRAS.447.3909R}. The S-type AGBs are AGBs that have rich s-elements like cAGBs but C/O less than unity. The super-AGB candidate MSX SMC055 is marked out.\label{fig:CMD}}
\end{figure*}

\begin{figure*}[ht!]
	\centering
	\includegraphics[width=0.9\linewidth]{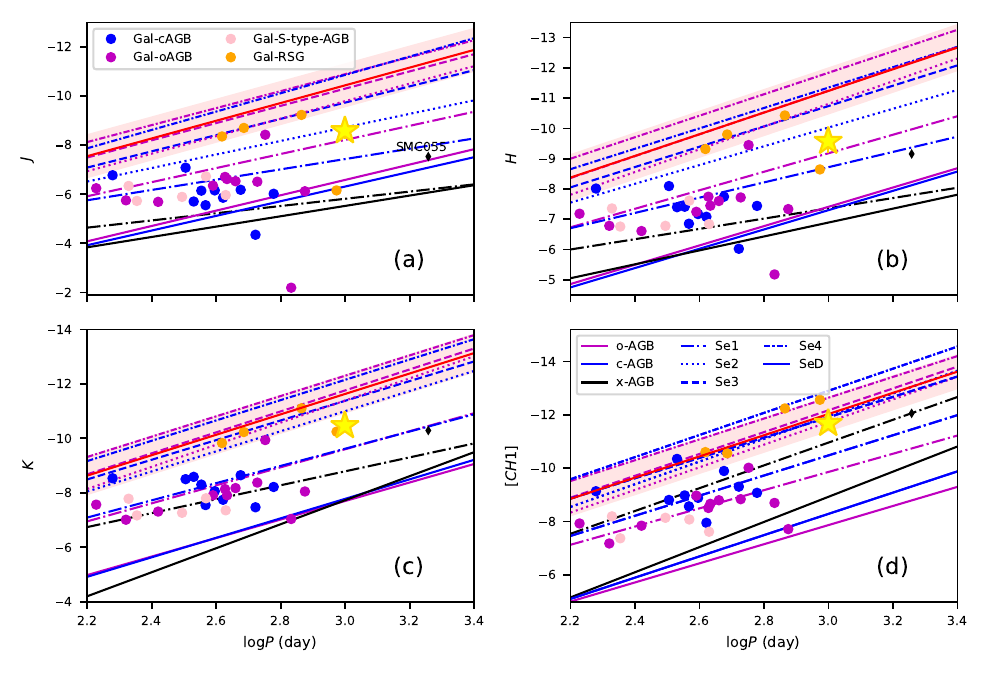}
	\caption{(a)$\sim$(d): Location of the progenitor of SN 2023ixf in the $P-L$ space compared with those derived for RSGs (red lines) \cite{2019ApJS..241...35R} in $J$, $H$, $K$ and $CH1$ bands, respectively. The light red region in each panel represents the 3-$\sigma$ range of the relation. LMC AGBs in different sequences \cite{2010ApJ...723.1195R} are plotted with different line styles and the colors denote different stellar types. Also plotted as filled dots are Galactic AGB and supergiants \cite{2015MNRAS.447.3909R}.\label{fig:P-L}}
\end{figure*}

To further examine the progenitor properties of SN 2023ixf, we compare its absolute magnitudes and colors with those of evolved massive star samples \cite{2009AJ....138.1003B,2015MNRAS.447.3909R} in Figure \ref{fig:CMD}. We find that the progenitor of SN 2023ixf has very red colors (i.e., $J-K\approx1.9$~mag, $J-[CH1]\approx3.2$~mag) compared with RSGs but are similar to AGB stars, though it appears more luminous than the latter. Only a few RSGs and possible super-AGB stars sit around it. 
Besides, with log$P\sim$3, the absolute $JHK$-band magnitudes inferred from the period-luminosity ($P-L$) relation \cite{2019ApJS..241...35R} are brighter than corresponding values of the progenitor by $>$1 magnitude, and the deviation decreases with wavelength, as shown in Figure \ref{fig:P-L}(a$\sim$c). 
However, the absolute magnitudes in $CH1$-band is in line with the $P-L$ relation of RSGs (see Figure \ref{fig:P-L}(d)).
Unlike RSGs, AGB stars have large dispersion in the $P-L$ space. This dispersion is partly contributed by their chemical types, i.e. carbon-rich (cAGB), oxygen-rich (oAGB), or highly reddened with indeterminate chemistry (xAGB).
But the dispersion is dominated by different evolutionary stages with different pulsation modes, which is evidenced by the five distinct sequences in the $P-L$ space (Se1--4 and SeD, denoted by different line styles in Figure \ref{fig:P-L}) \cite{1999IAUS..191..151W,2010ApJ...723.1195R}.
Inspecting Figure \ref{fig:P-L}, the progenitor of SN 2023ixf is located close to the relation of cAGBs in Sequence 2 (blue dotted lines in Figure \ref{fig:P-L}). But the long period of $\sim$1000 days is much longer than typical periods of Sequence 2 stars but comparable to Sequence D stars \cite{2010ApJ...723.1195R}. 
Thus, the progenitor seems to be a quite peculiar star.
Only very few supergiants, including the super-AGB candidate MSX SMC05, are found to show similar peculiar properties.

\subsection{Constraining properties of the progenitor and circumstellar dust}\label{sec:sed-fit}
The progenitor star of SN 2023ixf has similar properties to some extreme RSGs and super-AGB stars in the NIR/MIR bands.
These stars are also characterized by their dusty environment. 
Thus, the extreme red colors of the progenitor are likely due to large amount of dust around it.
Now we use the photometric results in Section \ref{sec:methods} to constrain the properties of the progenitor. 

\begin{figure}[H]
	\centering
	\includegraphics[width=1.0\linewidth]{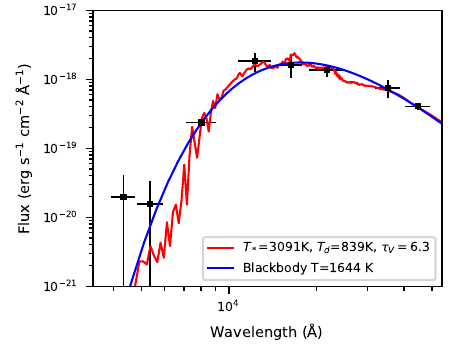}
	\caption{Spectral energy distribution of the progenitor of SN~2023ixf (black squares) and the best-fit DUSTY+MARCS model with log~g = $-$0.5, $R_{\mathrm{out}}/R_{\mathrm{in}}$ = 2.0. Also plotted is the best-fit blackbody model.\label{fig:sed_prog}}
\end{figure}

As shown in Figure~\ref{fig:sed_prog}, the SED of the progenitor can be well fit by a blackbody with $T_{\mathrm{eff}}\sim1644$~K, corresponding to a bolometric luminosity of log$(L/\mathrm{L_{\odot}})\sim$4.78 and a radius of $\sim3025$~R$_{\odot}$.
Such temperature is too low for RSGs (e.g. ref.~\cite{2012ApJ...750...97D} found $T_{\mathrm{eff}}\geq$3400~K for RSGs in M33).
As also discussed earlier in Section \ref{sec:spitzer-variation}, the progenitor is extremely red probably due to significant obscuration from dense circumstellar dust.

We use DUSTY, a 1-D code which solves the radiative transfer equation for a central source surrounded by a spherically symmetric dust shell at a certain optical depth \cite{1997MNRAS.287..799I}, to calculate the output flux of a dusty star.
The MARCS spectra models\footnote{\url{https://marcs.oreme.org/}} \cite{2008A&A...486..951G,2017A&A...601A..10V} with [Fe/H]=$-$0.25 are used as input for the external radiation source. 
DUSTY input parameters are the optical depth in $V-$band ($\tau_V$), temperature at the inner boundary $T_{\mathrm{d}}$ and the ratio of outer ($R_{\mathrm{out}}$) and inner boundaries ($R_{\mathrm{in}}$) of the dust shell. We adopt two sets of models with $R_{\mathrm{out}}/R_{\mathrm{in}}$=2.0, 10.0. Details of the fitting is presented in Appendix \ref{sec:sed-fit-method}.
The best-fit model with log~g = $-$0.5, $R_{\mathrm{out}}/R_{\mathrm{in}}$ = 2.0 has the minimum Chi-square of $\chi^2=0.570$. 
The best-fit model is shown in Figure \ref{fig:sed_prog} overlapped with the observed SED.
The resultant parameters for the progenitor of SN~2023ixf are listed in Table \ref{tab:fit-res}, which are $T_*$ = 3091$^{+422}_{-258}$~K, log$L/L_{\odot}=4.83^{+0.09}_{-0.03}$, $R_*$ = 912$^{+227}_{-222}$~R$_{\odot}$, $\tau_V$ = 6.25$^{+1.72}_{-0.85}$, and $T_d$ = 841$^{+351}_{-139}$~K. The outer radius of the dust shell is 25,400--117,200~\rsun (1.7$\times10^{15}$~cm--8.1$\times10^{15}$~cm).
Assuming a wind velocity of 70~km~s$^{-1}$, the mass loss rate is estimated as 6.22--9.41$\times10^{-6}$~\msun~yr$^{-1}$, and the total CSM mass is 0.6--3.0$\times10^{-4}$~M$_{\odot}$.
\begin{table*}[ht]
	\footnotesize
    \begin{threeparttable}
	\caption{Best-fit parameters for the DUSTY+MARCS models of the progenitor of SN~2023ixf. The mass loss rate is calculated by assuming a wind velocity of 70~km~s$^{-1}$. The lower and upper limits are given as the values at 16\%, 84\% of the posterior probability distribution of the MCMC sampling. The last column presents the values of $\chi^2$ of the best-fit model. }
	\label{tab:fit-res}
	\tabcolsep 8pt 
	\centering
	\begin{tabular*}{0.95\textwidth}{ccccccccccc}
	\toprule
	log~g &$R_{\mathrm{out}}/R_{\mathrm{in}}$ &$T_*$ &$T_d$ &$\tau_V$ &log($L$)\tnote{a)} &$R_{\mathrm{in}}$ &$R_*$ &$\dot{M}$ & $M_{\mathrm{w}}$ &$\chi^2$\\
		&        &(K)     &(K)     &        &(L$_{\odot}$)    &(10$^4$~R$_{\odot}$) &(R$_{\odot}$)   &(10$^{-6}$~\msun yr$^{-1}$) &(10$^{-5}$~M$_{\odot}$) & \\
	\hline
	$-$0.5 &2 &3091$^{3513}_{2833}$ &839$^{1003}_{564}$ &6.25$^{7.97}_{5.40}$ &4.83$^{4.92}_{4.80}$ &1.88$^{5.86}_{1.27}$ &912$^{1139}_{690}$ &6.97$^{9.41}_{6.22}$ &8.23$^{29.94}_{6.07}$ &0.570 \\
	0.0 &2 &3212$^{4383}_{3001}$ &841$^{1192}_{702}$ &6.61$^{8.68}_{5.88}$ &4.83$^{4.87}_{4.77}$ &1.89$^{3.09}_{1.02}$ &842$^{1010}_{425}$ &7.31$^{11.06}_{6.74}$ &8.72$^{13.58}_{6.13}$ &0.605 \\
	$-$0.5 &10 &3107$^{3588}_{2871}$ &860$^{1058}_{547}$ &4.84$^{6.25}_{4.26}$ &4.85$^{4.96}_{4.81}$ &1.75$^{6.62}_{1.09}$ &921$^{1114}_{694}$ &7.94$^{11.03}_{7.12}$ &43.70$^{215.23}_{29.14}$ &0.597 \\
	0.0 &10 &3245$^{4548}_{3109}$ &889$^{1300}_{770}$ &5.25$^{6.64}_{4.90}$ &4.84$^{4.87}_{4.78}$ &1.65$^{2.38}_{0.81}$ &837$^{972}_{400}$ &8.47$^{12.54}_{8.32}$ &43.91$^{61.94}_{27.84}$ &0.621 \\
	\bottomrule
	\end{tabular*}
    \begin{tablenotes}
    \item[a)] The uncertainties do not include the error in distance, which will add an additional uncertainty of 0.04.
    \end{tablenotes}
    \end{threeparttable}
\end{table*}

\subsection{Initial mass and mass loss history of the progenitor star} \label{sec:properties}
\begin{figure}[H]
	\centering
	\includegraphics[width=1.0\linewidth]{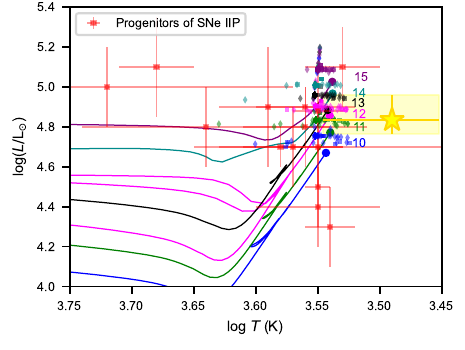}
	\caption{Location of the progenitor of SN~2023ixf (golden star) in the Hertzsprung Russell diagram overlapped the stellar evolution tracks of MIST single star models with [Fe/H]=$-$0.25, $v/v_{\mathrm{crit}}$=0.4 (colored solid lines). The filled dots represent the endpoints of the MIST models. Endpoints of the BPASS binary evolution models are plotted as empty diamonds (primary) and squares (secondary). Initial masses (in solar mass) are marked as the same color as the corresponding models. Also plotted are observed progenitors of SNe IIP \cite{2015PASA...32...16S,2019MNRAS.485.1990R}. \label{fig:HRD-prog}}
\end{figure}

In Section \ref{sec:sed-fit} we obtained the properties of the progenitor of SN~2023ixf by fitting its SED to spectral models of dusty stars. The progenitor may be an RSG surrounded by a condensed dust shell. We compare these results with the MESA Isochrones and Stellar Tracks (MIST) \cite{2016ApJ...823..102C,2016ApJS..222....8D} in Figure \ref{fig:HRD-prog}. The stellar tracks MIST library\footnote{\url{http://waps.cfa.harvard.edu/MIST/}} are computed with the Modules for Experiments in Stellar Astrophysics (MESA) code \cite{2011ApJS..192....3P,2013ApJS..208....4P,2015ApJS..220...15P,2018ApJS..234...34P}. It has large grids of single-star evolutionary models extending across all evolutionary phases for all relevant masses and metallicities. 
We select those with sub-solar metallicity ([Fe/H] = $-$ 0.25) and initial rotation rate of $v/v_{\mathrm{crit}}$ = 0.4. Figure~\ref{fig:HRD-prog} indicates that the progenitor of SN~2023ixf has a luminosity in range of other observed SNe~IIP progenitors and has an initial mass of 12$^{+2}_{-1}$~M$_{\odot}$. However, it may be the coolest SN progenitor ever discovered.

The progenitor of SN 2023ixf has extraordinary mass loss history as well as longer pulsational period among RSGs, which makes its observed properties in NIR bands not in line with the RSG family but in clusters of AGB stars and some rare extreme RSGs.
Mass loss in AGB stars can be enhanced by their strong pulsation and also dust formation \cite{2000ARA&A..38..573W}.
Super-AGB stars can be progenitors of some subluminous SNe II with low energy as well as low mass of sythesized $^{56}$Ni (e.g. SN 2018zd \cite{2021NatAs...5..903H}).
Observation of the SN can also give clues on its progenitor.
The $B$- and $V$-band light curves of SN 2023ixf highly resembles those of SN 2013by \cite{2015MNRAS.448.2608V}, with a plateau of $\sim$70 days  \cite{2023arXiv231010727Z}. The radioactive tail indicates an ejected \Ni mass of $\sim0.07$~M$_{\odot}$, which is much higher than expected by ECSNe ($\lesssim0.01$~\msun \cite{2006A&A...450..345K}).
Light curve fitting to hydrodynamic models suggested an initial mass of $\sim$13~\msun for SN 2013by \cite{2018ApJ...858...15M}.
We may assume that SN 2023ixf has a similar progenitor to SN 2013by. Meanwhile, progenitor of SN 2023ixf is located at the ``kink'' region in the $\dot{M}-L$ diagram\cite{2023A&A...676A..84Y}. With log$\dot{M}\sim-5.1$ and log$L$/\lsun$\sim$4.83, the final mass of the progenitor star is about 11.2~\msun using the relationship between $\dot{M}$, $L$ and stellar mass \cite{2023arXiv230908657V}.
To conclude, the progenitor star of SN 2023ixf was unlikely to be an sAGB star, whose initial mass is usually determined to be not higher than 11~M$_{\odot}$.

\begin{figure}[H]
	\centering
	\includegraphics[width=1.0\linewidth]{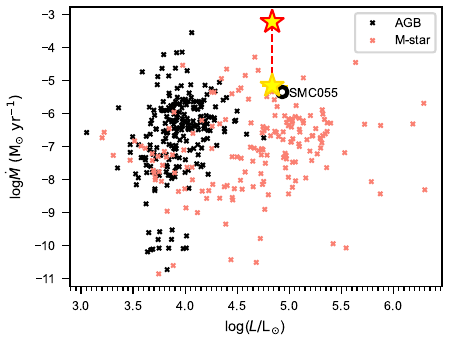}
	\caption{Relation between the luminosity and mass loss rate of the progenitor of SN~2023ixf derived from pre-explosion SED (golden star) and SN observations (red star), compared with AGB and M-type supergiant samples \cite{2018A&A...609A.114G}. The super-AGB candidate in the Small Magellanic Cloud MSX SMC055 is marked out as a black circle.\label{fig:logL-Mdot}}
\end{figure}

In Figure \ref{fig:logL-Mdot} we compare the mass loss rates and luminosities of M-type supergiants, AGB stars and the progenitor of SN~2023ixf.
The progenitor of SN~2023ixf has relatively high mass loss rate compared with both samples ($>$95\%(93\%) of the M-star(AGB) sample), but similar to extreme RSGs and sAGBs (e.g. MSX SMC 055).
Note that the wind velocity is assumed to be 70~km~s$^{-1}$, which is derived from the narrow emission lines in the SN spectra, but the wind velocity can be lower for the progenitor at $\sim$20 years before explosion thus the mass loss rate can be higher.
Analysis of the early time light curves and spectra yield a higher mass loss rate (e.g. $\sim6\times$10$^{-4}$~\msun yr$^{-1}$ \cite{2023arXiv230901998Z} or even $>$0.1~\msun yr$^{-1}$ \cite{2023ApJ...955L...8H}). 
It was proposed that the circumstellar matter was confined within a small radius ($R<10^{15}$~cm) around the exploding star, roughly consistent with our result.
This suggests that, during the final two decades on its way to the final core-collapse, the mass loss rate of the progenitor increased significantly.
As a result, the progenitor might have stripped off a small part of its envelope, shrinking in radius and getting hotter before explosion. 

The progenitor star of SN 2023ixf is identified as an RSG, but it exhibits unusual properties in the NIR $P-L$ diagram. This is probably attributed to extreme mass loss, stronger pulsation (longer period), and large amount of dust formation.
These factors are not independent.
The $P-L$ relation of RSGs may be related to several possible variables: initial mass, metallicity, mass loss rate, and rotation.
To test how these parameters affect the $P-L$ relation of RSGs, we make use of MESA, coupling with the non-adiabatic pulsation software \texttt{GYRE} \cite{2013MNRAS.435.3406T} to evolve a massive star and calculate the pulsation period until core carbon depletion.
The results show that the stellar luminosity decreases with larger metallicity, mixing length parameter or mass loss rate. While larger rotational velocity does not change the period.
However, metallicity is fixed for a star, and all stars with similar masses are assumed to share the same mixing length scale since they evolve under the same physical mechanisms.
Therefore, the excessively high mass loss rate may be the main cause of the downward shift of the progenitor in the $P-L$ diagram.
On the other hand, strong mass loss leads to increased dust formation around the star, which in turn contributes to stronger stellar winds.

The problem that remains is what causes such extreme winds in the progenitor of SN 2023ixf.
Binary interaction can be the answer.
In a binary system, mass loss of the progenitor might have been enhanced through tides, Roche lobe overflow or common-envelope evolution \cite{1988MNRAS.231..823T,2013Sci...339..433I,2007A&A...467.1181D}. 
Moreover, spectroscopic polarization observations reveal a high polarization of $\sim$1\% in the very early phases (day +2.5) and a quick drop to $\sim$0.5\% within one day. The correlation of the evolution of the flash emission lines in the SN spectra and the change of polarization implies that the CSM around the progenitor of SN 2023ixf is highly asymmetric \cite{2023ApJ...955L..37V}.
Relatively high polarization level later also suggests an aspherical SN ejecta.
The extraordinary polarization level in the early phases of SN 2023ixf is only seen two SNe IIP (SN 2013ej \cite{2021MNRAS.505.3664N}, SN 2021yja \cite{2023arXiv230306497V}) and some 87A-like SNe (SN 1987A \cite{1991ApJS...77..405J}, SN 2018hna \cite{2021MNRAS.503..312M,2021NatAs...5..544T}).
Thus, we suggest that the highly asymmetric and thick CSM is most likely to be produced through interaction with a companion star.

In Figure \ref{fig:HRD-prog}, we also show the endpoints of some binary stellar evolution models from the Binary Population and Spectral Synthesis code (BPASS) \cite{2017PASA...34...58E,2018MNRAS.479...75S}. Only models that may produce SNe II are plotted (i.e. with surface hydrogen abundance $X_\mathrm{H}>$0.5). 
With binary interaction, the models expand to a wide range in temperature. 
We consider two scenarios where the progenitor of SN 2023ixf was either the primary or the secondary star.
In the former case, where the progenitor star of SN 2023ixf is more massive, the efficient models cover all mass ratios ($q = M_1/M_2$) but only the most separated systems (with orbital period log$P\geq2.6$). The close system with log$P$=0.4 and a small mass ratio of 0.1, also lies in the range of the parameter space of the progenitor of SN 2023ixf. However, in this system, the two stars are separated only by about 7~R$_{\odot}$, which is far less than the stellar radius of the primary star thus is unlikely. 
In the latter case, where the progenitor star of SN 2023ixf is less massive, only long orbital systems are possible.
Both pictures suggest a binary system with long orbital periods.
It is not clear whether the companion is still alive, and can only be identified by future inspection of the SN site when the SN light fades away.


\section{Conclusion}
The progenitor of the very nearby type IIP supernova (SN) 2023ixf is studied with the pre-explosion images in optical to mid-infrared (MIR) bands. The progenitor candidate is resolved at the SN site in the multi-band \textit{HST} archive images, especially in the red bands. A point source was also detected at the SN location in the \textit{Spitzer}/IRAC $CH1$- and $CH2$-band images taken during period from 2004 to 2019. Photometry was performed for the progenitor candidate on the pre-discovery \textit{HST} and \textit{Spitzer} images. 

With additional near-and mid- infrared data, we find that the spectral energy distribution of the progenitor of SN 2023ixf shows a severe suppression in optical bands, implying a very low temperature and/or heavy extinction by the surrounding dust. In $CH1$- and $CH2$-bands, the progenitor star displays a periodic variability with a long period of 1033 days. The NIR and MIR colors, absolute magnitudes and the relation with the pulsation period of the progenitor are not in the range of normal RSGs or AGB stars but similar to some extreme RSGs and super-AGB stars, suggestive of the presence of significant dust around it. 

We constructed stellar spectral models with dust shell to fit the SED of the progenitor using the MARCS spectral library and the DUSTY code. 
The MCMC fitting gives a best-fit model with a relatively low stellar temperature of $T_*$ = 3091$^{+422}_{-258}$~K, which is the lowest ever known for SNe IIP progenitors. This low temperature, together with the derived luminosity log$(L/L_{\odot})=4.83^{+0.13}_{-0.07}$, suggests that SN 2023ixf has a dusty progenitor with initial masses of 12$^{+2}_{-1}$~M$_{\odot}$. 

Our spectral models indicate that the mass loss rate of the progenitor of SN 2023ixf is 6.22--9.41$\times10^{-6}$~\msun~yr$^{-1}$ at about 21 years before explosion, similar to that of the extreme RSGs with violent mass loss. On the other hand, the mass loss rate estimated from the early-time flash spectra ($\sim6\times10^{-4}$~\msun~yr$^{-1}$) is much higher than our result. This discrepancy indicates that the mass loss rate of the progenitor has increased significantly over the past two decades towards its final explosion. Combined with the nature of the SN itself, the progenitor star was unlikely to be a super-AGB, and the mechanism of enhanced mass loss is also different. Taking into account recent polarimetric results at early phases of SN 2023ixf, we proposed that the extraordinary mass loss of its progenitor may be a result of binary interaction. Compared with BPASS binary stellar models, the binary system of the progenitor of SN 2023ixf likely had a long period.

The unique properties of the progenitor of SN~2023ixf in NIR/MIR bands suggest that it was an extreme RSG with enhanced mass loss which was most likely due to binary interaction.
With late time observations of the SN, e.g. nebular phase spectra, the evolution of its progenitor can be better constrained.

\Acknowledgements{This work is supported by the National Natural Science Foundation of China (NSFC grants 12288102, 12033003, and 11633002), the Ma Huateng Foundation, the Scholar Program of Beijing Academy of Science and Technology (DZ:BS202002), and the Tencent Xplorer Prize. L.W. is sponsored (in part) by the Chinese Academy of Sciences (CAS), through a grant to the CAS South America Center for Astronomy (CASSACA) in Santiago, Chile. We acknowledge the support of the staff of the LJT. Funding for the LJT has been provided by the CAS and the People's Government of Yunnan Province. The LJT is jointly operated and administrated by YNAO and the Center for Astronomical Mega-Science, CAS.
This work made use of v2.2.1 of the Binary Population and Spectral Synthesis code (BPASS) models as described in refs.~\cite{2017PASA...34...58E,2018MNRAS.479...75S}.}

\InterestConflict{The authors declare that they have no conflict of interest.}



\bibliography{ref}
\bibliographystyle{scpma}

\begin{appendix}




\renewcommand{\thesection}{Appendix}

\section{}\label{sec:data}
\subsection{Pre-explosion \textit{HST} images and data reduction} \label{sec:HST}
We searched the pre-explosion \textit{HST} images from Mikulski Archive for Space Telescopes (MAST)\footnote{\url{http://archive.stsci.edu/}} and the Hubble Legacy Archive (HLA)\footnote{\url{http://hla.stsci.edu/}}, and found publicly available images in various bands taken from 1999 to 2018.

There is clearly a point-like source near the SN position in both the F814W and F658N images, but very faint in the others.
To get accurate positions of \mbox{SN~2023ixf} on the pre-discovery image, we made use of a drizzled ACS/WFC F814W image achieved from HLA as a pre-explosion image, and an image combined from 3 unfiltered 3-second images obtained by the 2.4-m Lijiang Telescope (LJT) on May, 20th, 2023 as a post-explosion image. We first chose 10 common stars that appeared on the LJT and \textit{HST} images and then got their positions on each image using \texttt{SExtractor}.
A second-order polynomial geometric transformation function is applied using the IRAF geomap task to convert their coordinates on the post-explosion image to those on the pre-explosion images.
Based on the \texttt{IRAF}\footnote{IRAF is distributed by the National Optical Astronomy Observatories, which were operated by the Association of Universities for Research in Astronomy, Inc., under cooperative agreement with the National Science Foundation (NSF).} \texttt{geoxytran} task, this established the transformation relationship between the coordinates of SN~2023ixf on the post-explosion image and those on the pre-explosion images.
The uncertainties of the transformed coordinates are a combination of the uncertainties in the SN position and the geometric transformation. The position of the progenitor candidate and SN~2023ixf in the pre-explosion images is shown in Figure~\ref{fig:hstimg}.
The locations of the progenitor on the other band images are obtained by either visually matching the images to the F814W image or transforming from the F814W image using similar coordinates transform procedures.

We use DOLPHOT\footnote{\url{http://americano.dolphinsim.com/dolphot/}} 2.0 to get photometry of the progenitor on the pre-explosion images with parameters recommended in its \emph{User's Guide}.
The photometry is performed on the bias-subtracted, flat-corrected, so-called C0M FITS images for WFPC2/WFC instrument, and bias-, flat-, CTE-corrected, so-called FLC FITS images for ACS/WFC and ACS/WFC instruments, all obtained from the \mbox{MAST} archive.
Choosing the F814W image as a reference image, DOLPHOT is run simultaneously on multiple-band images taken on the same day.
Magnitudes and their uncertainties of the progenitor candidate are extracted from the output of DOLPHOT. The photometry results are listed in Table~\ref{tab:hst-obs}.

\subsection{Pre-explosion images from the \textit{Spitzer} Infrared Array Camera (IRAC)} \label{sec:Spitzer}
The SN~2023ixf field in M101 was observed with the \textit{Spitzer}/IRAC before its explosion by several programs from 2004 to 2019 by PI G. Rieke with program ID 60, by PI M. Kasliwal with program IDs 10136, 11063, 13053, 14089, 80196, and 90240, and by PI P. Garnavich with program ID 80126. We utilized the level 2 post-BCD (Basic Calibrated Data) images from the \textit{Spitzer} Heritage Archive (SHA)\footnote{\url{http://irsa.ipac.caltech.edu/applications/Spitzer/SHA/}}, which were reduced by the {\it Spitzer} pipeline and resampled onto $0.6''$ pixels. A point source is detected with $2\sigma$ detection threshold at $CH1$ ($3.6\mu$m) and $CH2$ ($4.5\mu$m) from 2004 to 2019, while there is no detection in $CH3$ ($5.8\mu$m) and $CH4$ ($8.0\mu$m) bands in the year 2004. 

Aperture photometry was performed on the pre-explosion images of the SN field with an aperture radius of 4 pixels (2.4 arcsecs). Aperture corrections were applied following the IRAC Data Handbook. 
The level 2 post-BCD images have been calibrated in an absolute surface-brightness unit of MJy/sr, which can be transformed into units of $\mu$Jy/pixel$^2$ by a conversion factor of 8.4616 for the angular resolution of the IRAC images $0.6''$ pixels. The flux was converted to AB magnitude according to the definition $m_{\mathrm{AB}}=-$2.5log10($f$)+23.9, where $f$ is in units of $\mu$Jy \cite{Fukugita96}. 
The AB magnitudes and fluxes of the SN field at $CH1$ and $CH2$ bands are listed in Table~\ref{tab:spitzer-phot}.

\subsection{SED fit method}\label{sec:sed-fit-method}
DUSTY has been widely used to model the SEDs of stars and estimate their mass loss rates \cite{2012ApJ...759...20K,2018MNRAS.481.2536K,2021ApJ...912..112W,2020ApJ...891...43S}.
We assume a dust composition of 100\% graphite \cite{1984ApJ...285...89D}, as ionized carbon lines emerge in the early supernova flash.
The dust grain size follows the standard MRN power-law \cite{1977ApJ...217..425M} (i.e. $n(a)\propto a^{-3.5}$, for $0.005\leq a \leq 0.25~\mathrm{\mu}$m).
The MARCS spectra models \cite{2008A&A...486..951G,2017A&A...601A..10V} are used as input for the external radiation source. The MARCS models have standard composition and spherical geometry. We select models with [Fe/H]=$-$0.25. The effective temperatures are $T_*$ = 2600--8000~K for log~g = 0.0, 1.0, 2.0, 3.0, while $T_*$ = 2500--3900~K for log~g = $-$0.5. DUSTY input parameters are the optical depth in $V-$band ($\tau_V$), temperature at the inner boundary $T_{\mathrm{d}}$ and the ratio of outer and inner boundaries of the dust shell. We adopt two sets of models with $R_{\mathrm{out}}/R_{\mathrm{in}}$=2.0, 10.0, like in earlier studies \cite{2012ApJ...759...20K}.
DUSTY produces the relative flux ($f_{\lam}=\lam F(\lam)/\int F(\lam)\mathrm{d}\lam$), and the flux at the inner boundary ($F_1$) is normalized to a total luminosity of $10^4~L_{\odot}$, for each model.
The DUSTY output spectra are self-similar so can be rescaled to any required luminosity. 
Thus, the total relative flux is dependent on $T_*$, $T_d$, $\tau_V$, and dust shell radius is scaled as $(L/10^4\mathrm{L_{\odot}})^{1/2}$.

We refer to the method of previous studies \cite{2018MNRAS.481.2536K,2021ApJ...912..112W} to fit the observed SED of the progenitor.
The best-fit model is determined by minimizing the chi-square as below:
\begin{equation}
	\label{equ:chi2}
	\chi^2 = \sum^{N}\frac{1}{N-p-1}\frac{[f(O, \lam)-f(M, \lam)]^2}{\sigma(O, \lam)^2}
\end{equation}
in which $f(O, \lam) = \frac{F(O, \lam)}{F(O, K)}$, $f(M, \lam) = \frac{F(M, \lam)}{F(M, K)}$ are the observed (O) and model (M) flux normalized to the $K$-band, respectively; $\sigma(O, \lam)$ is the uncertainty of the observed flux divided by $f(O, K)$; $N$, $p$ are the number of data points and number of free parameters, respectively ($N=8$, $p=3$).
According to the scaling relation of DUSTY, the wind mass loss rate (including gas and dust and assume a gas-to-dust ratio of 200) is determined as below:
\begin{equation}
	\dot{M} = \dot{M}_{DUSTY}\frac{L}{10^4\mathrm{L}_{\odot}}(\frac{v_\mathrm{w}}{v_{DUSTY}})^{-1}
\end{equation}
and the total CSM mass is $M_\mathrm{w} = \dot{M} R_{\mathrm{out}}/v_\mathrm{w}$.

A Markov-chain Monte Carlo \texttt{python} package \texttt{emcee} \cite{2013PASP..125..306F} is applied to do the fitting. During the fitting, models with parameters not in our constructed model grids are calculated by linear interpolation between the girdded models.
We test the fit with different $R_{\mathrm{out}}/R_{\mathrm{in}}$s and log~gs.
With each $R_{\mathrm{out}}/R_{\mathrm{in}}$, we notice that with the resulted radii and log~gs, the star's mass would be extremely high ($>$200~M$_{\odot}$) except with log~g = $-$0.5 or 0.0. Thus results for log~gs $\geq$ 1.0 are dropped.
We list the results of our fitting for log~g = $-$0.5, 0.0 in Table \ref{tab:fit-res}.
\begin{table*}[ht]
	\footnotesize
	\caption{The $CH1$ and $CH2$ bands photometry on the Pre-explosion \textit{Spitzer}/IRAC images on the site of \mbox{SN~2023ixf}. The AB magnitudes are listed in this table. }
	\label{tab:spitzer-phot}
	\centering
	\tabcolsep 8pt 
	\begin{tabular*}{0.93\textwidth}{cccccccccc}
	\toprule
	Date &MJD &Flux$_{CH1}$ &$\sigma(\mathrm{flux})_{CH1}$ &AB$_{CH1}$ &$\sigma(\mathrm{AB})_{CH1}$ &Flux$_{CH2}$ &$\sigma(\mathrm{flux})_{CH2}$  &AB$_{CH2}$ &$\sigma(\mathrm{AB})_{CH2}$ \\
	(yy-mm-dd) & (days) & ( $\mu$Jy ) & ( $\mu$Jy ) &  ( mag ) & ( mag ) & ( $\mu$Jy ) & ( $\mu$Jy ) & ( mag ) & ( mag ) \\
	\hline
	2004-03-08 & 53072.09 & 25.48 & 8.35 & 20.38 & 0.36 & 21.67 & 7.44 & 20.56 & 0.37 \\
	2004-03-08 & 53072.49 & 30.41 & 8.56 & 20.19 & 0.31 & 29.02 & 7.99 & 20.24 & 0.30 \\
	2012-02-03 & 55960.72 & 25.55 & 7.61 & 20.59 & 0.36 &  ...  &  ...  &  ...  &  ...  \\
	2012-02-23 & 55980.99 & 21.15 & 7.11 & 20.38 & 0.32 &  ...  &  ...  &  ...  &  ...  \\
	2012-08-26 & 56165.01 &  ...  &  ...  &  ...  &  ...  & 20.96 & 6.82 & 20.60 & 0.35 \\
	2013-02-14 & 56337.07 & 16.01 & 6.72 & 20.50 & 0.35 &  ...  &  ...  &  ...  &  ...  \\
	2013-02-25 & 56348.11 & 23.01 & 7.37 & 20.89 & 0.46 &  ...  &  ...  &  ...  &  ...  \\
	2013-08-12 & 56516.35 &  ...  &  ...  &  ...  &  ...  & 17.72 & 6.42 & 20.78 & 0.39 \\
	2014-03-26 & 56742.84 & 26.50 & 7.88 & 20.30 & 0.32 & 28.83 & 7.80 & 20.25 & 0.29 \\
	2014-04-24 & 56771.83 & 36.62 & 8.94 & 19.99 & 0.27 & 28.00 & 7.87 & 20.28 & 0.31 \\
	2014-09-02 & 56902.01 & 27.47 & 8.01 & 20.34 & 0.32 & 31.68 & 8.34 & 20.15 & 0.29 \\
	2015-04-24 & 57136.69 & 18.61 & 6.85 & 20.73 & 0.40 & 18.52 & 6.41 & 20.73 & 0.38 \\
	2015-05-02 & 57144.06 & 18.27 & 6.98 & 20.75 & 0.41 & 16.86 & 6.48 & 20.83 & 0.42 \\
	2015-05-08 & 57150.17 & 20.12 & 7.19 & 20.64 & 0.39 & 14.27 & 5.87 & 21.01 & 0.45 \\
	2015-05-21 & 57163.71 & 19.22 & 7.20 & 20.69 & 0.41 & 27.98 & 7.60 & 20.28 & 0.29 \\
	2015-06-18 & 57191.82 & 9.21 & 5.83 & 21.49 & 0.69 & 10.41 & 5.23 & 21.36 & 0.55 \\
	2015-07-17 & 57220.79 & 17.08 & 6.74 & 20.82 & 0.43 & 13.49 & 5.70 & 21.07 & 0.46 \\
	2015-08-13 & 57247.82 & 16.28 & 6.44 & 20.87 & 0.43 & 13.46 & 5.57 & 21.08 & 0.45 \\
	2016-04-08 & 57486.85 & 29.31 & 8.17 & 20.23 & 0.30 & 24.16 & 7.13 & 20.44 & 0.32 \\
	2017-03-31 & 57843.93 & 38.08 & 8.91 & 19.95 & 0.25 & 31.36 & 8.14 & 20.16 & 0.28 \\
	2017-06-22 & 57926.90 & 33.59 & 8.68 & 20.08 & 0.28 & 24.78 & 7.33 & 20.41 & 0.32 \\
	2017-09-13 & 58009.67 & 23.86 & 7.44 & 20.46 & 0.34 & 20.59 & 6.78 & 20.62 & 0.36 \\
	2018-04-24 & 58232.95 & 24.43 & 7.47 & 20.43 & 0.33 & 20.58 & 6.70 & 20.62 & 0.35 \\
	2018-06-23 & 58292.87 & 18.90 & 6.86 & 20.71 & 0.39 & 15.15 & 6.21 & 20.95 & 0.44 \\
	2018-09-19 & 58380.22 & 15.84 & 6.61 & 20.90 & 0.45 & 11.69 & 5.47 & 21.23 & 0.51 \\
	2019-03-30 & 58572.08 & 33.48 & 8.42 & 20.09 & 0.27 & 24.00 & 7.27 & 20.45 & 0.33 \\
	2019-05-11 & 58614.39 & 27.15 & 7.55 & 20.32 & 0.30 & 23.58 & 7.21 & 20.47 & 0.33 \\
	2019-06-21 & 58655.68 & 28.80 & 8.09 & 20.25 & 0.30 & 27.84 & 7.69 & 20.29 & 0.30 \\
	2019-08-02 & 58697.50 & 43.15 & 9.47 & 19.81 & 0.24 & 25.69 & 7.51 & 20.38 & 0.32 \\
	2019-09-14 & 58740.01 & 20.04 & 7.22 & 20.65 & 0.39 & 21.97 & 7.04 & 20.55 & 0.35 \\
	2019-10-25 & 58781.31 & 30.37 & 8.56 & 20.19 & 0.31 & 30.46 & 8.11 & 20.19 & 0.29 \\
	\bottomrule
	\end{tabular*}
\end{table*}

\end{appendix}

\end{multicols}
\end{document}